\documentclass[twocolumn,showpacs,aps,prl]{revtex4}
\usepackage{graphicx}
\usepackage{amsmath}
\usepackage{amssymb}
\usepackage{units}

\newcommand{\aver}[1]{\langle #1 \rangle}

\newcommand{\mul}[2]{\lvert #1 \rangle\langle #2 \rvert}
\newcommand{\tr}[1]{\ensuremath{\mathrm{Tr}\{#1\}}}
\newcommand{\av}[1]{\overline{#1}}
\newcommand{\ellel}{\ell_{\mathrm{el}}}

\begin{document}

\title{Anomalous diffusion and elastic mean free path in disorder-free 
  multi-walled carbon nanotubes}

\author{Shidong Wang and Milena Grifoni}
\affiliation{Theoretische Physik, Universit\"at Regensburg, 93040 Germany}
\author{Stephan Roche } 
\affiliation{CEA/DSM/DRFMC/SPSMS/GT, 17 avenue des Martyrs,
38054 Grenoble, France}
\date{\today}

\begin{abstract}
We explore the nature of anomalous diffusion of wave packets in
disorder-free incommensurate multi-walled carbon nanotubes. The
spectrum-averaged diffusion exponent is obtained by calculating the
multifractal dimension of the energy spectrum. Depending on the shell
chirality, the exponent is found to lie within the range $1/2 \leq \eta <
1$. For large unit cell mismatch between incommensurate shells, $\eta$
approaches the value 1/2 for diffusive motion. The energy-dependent
quantum spreading reveals a complex density-of-states-dependent
pattern with ballistic, super-diffusive or diffusive character.
\end{abstract} 

\maketitle



The understanding of charge transport in structurally clean systems,
with complex and aperiodic long-range correlations, has been the
subject of intense debate during the past two decades~\cite{QP-INC}.
The quantum dynamics in most of these systems has been described as an
anomalous quantum diffusion process, related to the multifractal
nature of the electronic states and spectra~\cite{sokoloff:pr1985}.
These unconventional transport mechanisms are incompatible with a
dominant transport length scale such as the elastic mean free path,
and the occurrence of a diffusive regime. Recently, the discovery of
carbon nanotubes has provided a whole class of new
quasi-one-dimensional systems with spectacular effects of topological
arrangements of carbon atoms on the electronic
spectra~\cite{saito93,Saito}. The multi-walled nanotubes (MWNTs) are
intrinsic incommensurate objects since, due to registry mismatch
between neighboring shells, there are very few cases in which the
respective symmetries of individual shells allow finding a common unit
cell for the whole object. In most cases, the unit cell length (along
the nanotube axis) ratio between adjacent shells is an irrational
number, and the MWNT taken as a whole becomes an incommensurate
object~\cite{saito93}.  In recent years, mechanical~\cite{Kolmogorov}
and electronic~\cite{Roche2001, Yoon2002, Ahn2003, Triozon2004, dwnt,
  Lunde2005,Wang2005} properties of double- and triple-walled
incommensurate nanotubes (i-DWNTs and i-TWNTs, respectively) have been
intensively
investigated. Roche \emph{et al.}~\cite{Roche2001} first reported on
anomalous diffusion properties of i-DWNTs and 
i-TWNTs by numerical analysis of the wave-packet propagation.
Signatures of anomalous diffusion were further inferred by the
evaluation of the energy spacing distribution of energy
levels~\cite{Ahn2003}, exhibiting Wigner-Dyson, Poisson and
semi-Poissonian statistics depending on the position of the Fermi
level. In a recent work~\cite{Wang2005}, the existence of a finite
(energy-dependent) elastic scattering rate for electrons in the outer
shell of a 
disorder-free i-DWNTs was analytically shown to be a consequence of
helicity-determined selection rules for inter-shell
tunneling~\cite{Yoon2002,Lunde2005}.

Aim of this letter, is to establish the connection between spectral and
clarify the dynamical properties of i-DWNTs, i-TWNTs and in general
MWNTs. Our results provide a deeper insight on why, experimentally,
MWNTs typically, but not
always~\cite{Frank1998,Urbina2003} exhibit normal diffusion behavior,
with an energy dependent mean free path~\cite{Stojetz2005}. However,
we predict global anomalous diffusive behavior in i-DWNTs and in many
i-TWNT systems.
To this extend, first we calculate the \emph{spectrum-averaged}
diffusion exponent $\eta$, which describes 
the spread in time of an initially localized wave packet by looking at
the mean square displacement
\begin{equation}
\label{eq:exponent}
  \av{X^2} (t) \simeq t^{2 \eta} \;.
\end{equation}
For this we evaluate the
multifractal dimension $1/2 \leq D_{-1} <1$ of the energy spectrum
of various incommensurate, tunnel coupled, i-DWNTs and i-TWNTs,
and use the relation $\eta =
D_{-1}$~\cite{Piechon1996,Ketzmerick1997}. 
We find a good agreement with previous calculations using a 
wave-packet propagation approach~\cite{Roche2001,Triozon2004}. 
We find that the exponent $\eta$
strongly depends on the chirality and on the number of shells. In
particular, for fixed number of
shells, e.g. two, one can find that $\eta$ is closer to the value
1/2, characteristic of normal diffusion,
if at least one shell has a large unit cell. Also upon increasing
the number of shells the diffusive limit 1/2 is more rapidly approached. For
example, for the i-TWNT $(6,4)@(17,0)@(15,15)$ is already
$\eta\simeq 1/2$ such that for such a tube some elastic mean
free path can be extracted from the wave-packet evolution. In
contrast, for the $(5,5)@(17,0)@(15,15)$ i-TWNT, with a very small
unit cell of the (5,5) shell, the diffusion remains anomalous,
and thus no \emph{global} mean free path can be extracted. While the exponent
$\eta$ describes spectrum-averaged properties, more information about
the interplay between density of states and degree of anomaly can be
extracted from the energy dependent quantities. Thus
wave-packet spreading is investigated by solving the time-dependent 
Schr\"odinger equation and calculating the \emph{energy-dependent} quantum
spreading
\begin{equation*}
\ell(E,t) =  \frac{X^{2}(E,t)}{v(E)t} =
\frac{\tr{\delta(E-\hat{\mathcal{H}}) 
      (\hat{\mathcal{X}}(t)-\hat{\mathcal{X}}(0))^2}}
      {\tr{\delta(E-\hat{\mathcal{H}})} \, v(E)t } \;.
\end{equation*}
Here $\delta(E-\hat{\mathcal{H}})$ is the spectral measure operator,
whose trace gives the total density of states $n(E)$,
$\hat{\mathcal{H}}$ is the system's Hamiltonian, and
$\hat{\mathcal{X}}(t)$ is the position operator along the tube
axis. Finally, $v(E)$ is the group velocity at energy $E$.
In some situations $\ell(E,t)$ becomes
time-independent such that energy-dependent elastic
mean free paths can be defined even if the global exponent $\eta$ is not
$1/2$!

\begin{figure}[htbp]
  \begin{center}
    \includegraphics[width=1.0\linewidth]{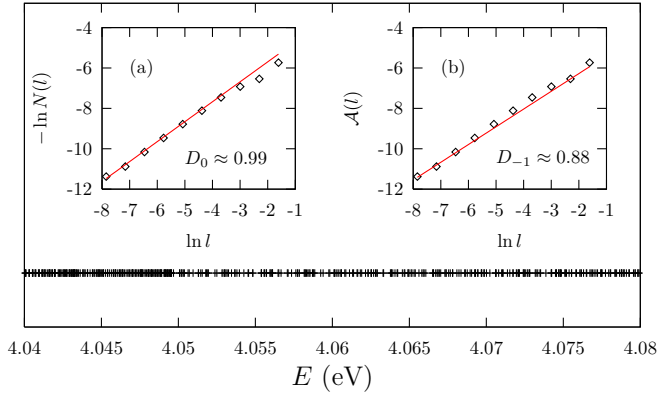}

  \end{center}
  \caption{(Color online) Fractal character of the energy spectrum of
    the i-DWNT (9,0)@(10,10) in the range
    $[\unit[4.04]{eV},\unit[4.08]{eV}]$.  Inset (a): 
     Plot of $-\ln N(l)$ vs. $\ln l$ as determined from the
    energy spectrum. A linear fit to the data corresponding to the
    shortest lengths $l$ yields the fractal dimension $D_{0}$. 
    Inset (b): Plot of ${\cal A}_l=-\frac{1}{2}\ln
    \sum_{i=1}^{N(l)} p_i^{-1}(l)$ vs. $\ln l$ as determined from the
    energy spectrum. The slope is the fractal dimension
     $D_{-1}$.  }
  \label{fig:DWNT}
\end{figure}

Our starting model to evaluate wave-packet spreading is the
tight-binding Hamiltonian for a MWNT with $M$ shells, where only
one $p_{\perp}$-orbital per carbon atom is kept, and with zero
on-site energies. With a nearest-neighbor hopping $\gamma_0$ on
each layer $n$, and hopping $\beta$ between neighboring layers, it
has the form~\cite{Lambin2000}:
\begin{equation}
  \label{eq:hamiltonian}
  \begin{split}
    \hat{\cal H} &= \gamma_{0} \sum_{n=1}^M\sum_{\aver{i,j}}
    \mul{p_{\perp}^{n,j}}{p_{\perp}^{n,i}} \\
    &\quad - \beta \sum_{\aver{n,m}}\sum_{i,j} \cos(\theta_{ij}) \,
    {e}^{-\frac{d_{ij}-a}{\delta}} \, 
    \mul{p_{\perp}^{j,n}}{p_{\perp}^{i,m}} \;,
  \end{split}
\end{equation}
where $\aver{n,m}$ and $\aver{i,j}$ are sums over nearest shells and
nearest neighboring atoms, respectively. Moreover, $\theta_{ij}$ is the
angle between the $p_{\perp}^{i}$ and $p_{\perp}^{j}$ orbitals, and $d_{ij}$
denotes their relative distance. The parameters used here are:
$\gamma_{0}=\unit[2.9]{eV}$, $a= \unit[3.34]{\AA}$, $\delta=
\unit[0.45]{\AA}$~\cite{saito93}.  An ab-initio estimate gives for the
intershell coupling $\beta\simeq\gamma_{0}/8$~\cite{Lambin2000}. Starting from
Eq.~\eqref{eq:hamiltonian}, the spreading of wave packets is now
evaluated along two different routes. For our first approach, we start
by observing that the dynamics of wave packets in a system is strongly
related to the properties of the system's energy spectrum.  The latter
can be generally divided into absolutely continuous, singular
continuous and pure point parts~\cite{last:jfa1996}. The wave packets
spread ballistically if the energy spectrum is absolute continuous and
are localized in systems with purely point-like spectra. If the
spectrum is singular continuous the wave packets anomalously
spread~\cite{guarneri, ketzmerick:prl1992, Geisel:prl1991}.  The
singular continuous spectrum is a multifractal object which can be
characterized by a set of fractal dimensions
$D_q$~\cite{paladin:pr1987}.  Pi\'{e}chon has shown that, at large
times $t \to \infty $, the moments spread as $\av{X^q}(t) \sim t^{q D_{1-q}}$,
with $D_q$ the fractal dimension of the energy
spectrum~\cite{Piechon1996}.  Then, the spreading of a wave packet is
determined by ${D_{-1}}.$ The motion of wave packets in a system with
singular continuous energy spectrum will be normal diffusive if
$D_{-1} = 1/2$, or anomalous diffusive if $1/2 < D_{-1} < 1$. The
energy spectrum of an incommensurate system is usually singular
continuous, or maintain intrinsic self-similar
features~\cite{sokoloff:pr1985}. Therefore, one may expect some
anomalous diffusive behavior of wave packets in i-DWNTs. In order to
calculate the energy spectrum of an i-DWNT, we approximate the
irrational ratio of the unit cell lengths of the inner and outer
shells by one of its convergents, which is the rational number
obtained by truncating the continued fraction representation of the
given irrational number up to a certain term~\cite{sokoloff:pr1985,
  goldman:rmp1993} That is, we approximate the i-DWNT by a
commensurate DWNT. As the chosen convergent becomes closer to the
irrational unit cell ratio, the energy spectrum of the commensurate
DWNT gets closer to that of the i-DWNT under study.  We calculate the
energy spectrum of the commensurate DWNT by direct diagonalization of
the Hamiltonian in Eq.\eqref{eq:hamiltonian}. After computing the
energy spectrum, we count the numbers of boxes $N(l)$ with length $l =
\Delta E/2^n$ needed to cover it. Here $\Delta E$ is the range of the energy
spectrum and $n$ is a positive integer. Thus, the probability density
$p_i(l)$ is obtained for each box by calculating the ratio of the
number of points falling into the box $\Lambda_i(l)$ to the total number of
points in the data. The set of fractal dimensions $D_q$ can be defined
as
\begin{equation*}
 \sum_{i=1}^{N(l)} p_i^{q+1}(l) \sim l^{q D_{q+1}} \;, \qquad \text{as } l \to
 0 \;.
\end{equation*}
For a simple fractal object, that is, the probability density
$p_i(l)$ is the same for each box $\Lambda_i(l)$, all general
dimensions defined above have the same values $D_F$. In this case,
the probability density is $p_i(l) = 1/N(l)$,
with $N(l) \sim l^{-D_F}$. Therefore is $D_{q+1} = D_F$ for
any $q$. Here, $D_F$ is the  fractal (box-counting) dimension.
In general, the distribution of points in a fractal object is
different for different boxes. That is, the probabilities of
finding a point in the different boxes are different.
If $q=-1$, one has $D_0 = D_F$.  For $q=-2$ the dimension
$D_{-1}$ can be calculated by
\begin{equation}
  D_{-1} = -\frac{1}{2} \lim_{l\to0} \frac{\ln \sum_{i=1}^{N(l)}
    p_i^{-1}(l)}{\ln l} \;.
\end{equation}
Hence, the dimension $D_{-1}$ is extracted from a linear regression
fit to the plot of ${\cal A}_l\equiv -\frac{1}{2}\ln\sum_{i=1}^{N(l)}
p_i^{-1}(l)$ as a function of $\ln l$.  As an example,
Fig.~\ref{fig:DWNT} shows results for the energy spectrum of the
i-DWNT $(9,0)@(10,10)$. As discussed in Ref.~\cite{Wang2005}, the
incommensurability of the two unit cells yields a non-vanishing
intershell tunneling only when helicity-determined selection rules are
fulfilled, which can occur only if enough sub-bands in each shell
become populated. In turn, this yields a finite lifetime for electrons
in one shell due to effective back-scattering processes.  Here we show
that these features are also reflected in the energy spectrum which,
as shown in Fig.~\ref{fig:DWNT}, exhibits a fractal character. We find
that within the energy range $[\unit[-12.664]{eV}, \unit[12.898]{eV}]$
the fractal dimension converges to the value $\eta=0.88$, when the
incommensurate ratio $T_{(10,10)}/T_{(9,0)}= 1/\sqrt{3}$ is
approximated by the 6th convergent $15/26$ obtained by truncating the
continued fraction expression of $1/\sqrt{3}$. This value is the same
as found from numerical wave-packet propagation of initially localized
wave packets~\cite{Roche2001}. Here,
$T_{(n,m)}=a_0\sqrt{3(n^2+nm+m^2)}/{\rm GCD}$ is the length of the
axial unit cell of shell $(n,m)$, with the carbon-carbon bond length
$a_0$ and GCD being the greatest common divisor of $(2n+m)$ and
$(m+2n)$.


\begin{figure}[htbp]
  \begin{center}
    \includegraphics[width=1.0\linewidth]{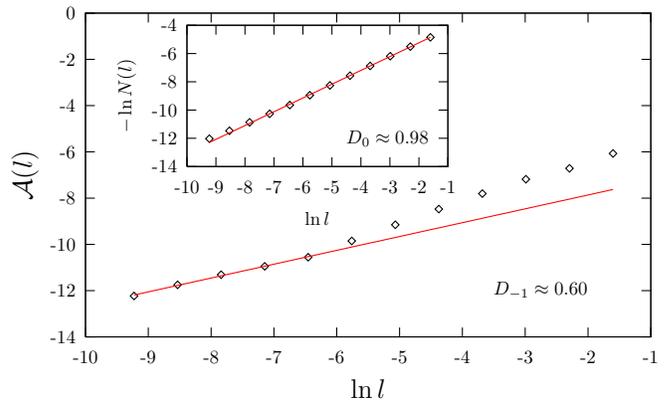}

  \end{center}
  \caption{(Color online) Plot of ${\cal A}_l=-\frac{1}{2}\ln
    \sum_{i=1}^{N(l)} p_i^{-1}(l)$ vs. $\ln l$ as extracted from the
    energy spectrum of the $(6,4)@(17,0)@(15,15)$ i-TWNT. The linear
    fit to the data corresponding to the shortest box lengths $l$
    yield the fractal dimension $D_{-1}$. Inset: Plot of $-\ln N(l)$
    vs. $\ln l$ for the same i-TWNT. The slope is the fractal
    dimension $D_0$.  } \label{fig:TWNT}
\end{figure}

In order to understand the role of chirality, we have also
investigated the cases of the $(6,4)@(17,0)$ and $(5,5)@(17,0)$
i-DWNTs, differing in the chirality of the inner shell. One has
$T_{(17,0)}/T_{(5,5)}=\sqrt{3}$, while
$T_{(17,0)}/T_{(6,4)}=1/\sqrt{19}$. The difference in the two
exponents is noticeable: $\eta_{(5,5)@(17,0)} \approx 0.88$ and
$\eta_{(6,4)@(17,0)} \approx 0.82$. We attribute the smaller exponent of the
$(6,4)@(17,0)$ i-DWNT compared to that of the $(5,5)@(17,0)$ i-DWNT to
the larger unit shell mismatch.
Having in mind the i-DWNT $(5,5)@(17,0)$, we add an additional
outer shell and consider  the i-TWNTs $(5,5)@(17,0)@(15,15)$ and
$(6,4)@(17,0)@(15,15)$. For the first TWNT the diffusion is still
anomalous with
 $\eta_{(5,5)@(17,0)@(15,15)} \approx 0.88$. However, the effect of the
additional armchair shell $(15,15)$ is to \emph{randomize} the energy
spectrum of the $(6,4)@(17,0)@(15,15)$. 
To be definite, already for the approximation $26:26:15$ to the ratios
$3\sqrt{19}:3:\sqrt{3}$ we find the value $\eta \approx 0.60$, as
shown in Fig.~\ref{fig:TWNT}. Notice that, since $D_{-1}$ is
defined in the limit of box length $l\to 0$, for the linear fit in
Fig.~\ref{fig:TWNT} the five points corresponding to the
smallest lengths $l$ have been considered. 

\begin{figure}[htbp]
  \begin{center}
    \includegraphics[width=\linewidth]{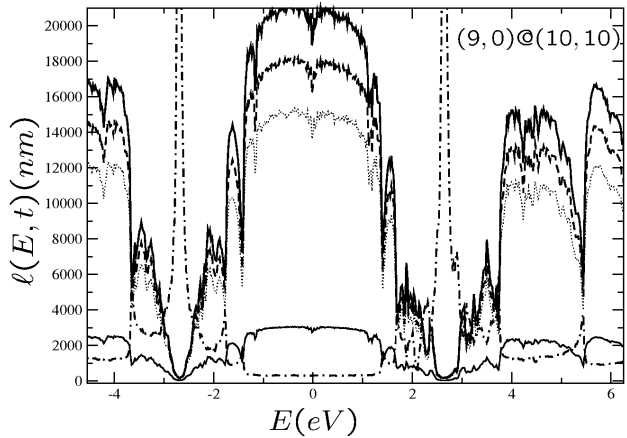}
  \end{center}
  \caption{(Color online) Energy-dependent behavior of the spreading
    $\ell(E,t)$ for the 
    $(9,0)@(10,10)$ i-DWNT, at several times $t$ (solid line
    $t=1000T$, dotted $t=5000T$, dashed $t=6000T$, bold $t=7000T$,
    with $T=14\hbar/\gamma_{0}$). The rescaled total density of states is also
    shown (dash dotted-line). 
  }
\label{fig:DWNT-wp}
\end{figure}

\begin{figure}[htbp]
  \begin{center}
\includegraphics[width=\linewidth]{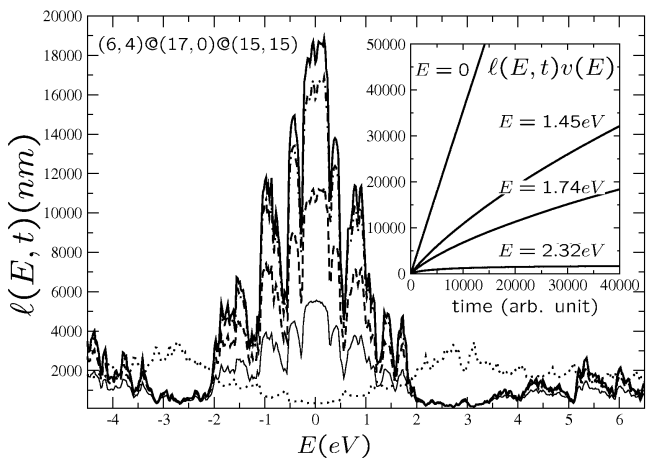}
  \end{center}
  \caption{(Color online) Energy-dependent behavior of $\ell(E,t)$ for the
    $(6,4)@(17,0)@(15,15)$ i-TWNT, at several times $t$ (solid line
    $t=2000T$, dashed $t=4000T$, dot-dashed $t=6000T$, bold $t=6710T$,
    with $T=14\hbar/\gamma_{0}$). The rescaled total density of states is also
    shown (dotted-line). Inset: time-dependence of the diffusivity
    $\ell(E,t)v(E)$ at various selected energies. 
  }
\label{fig:TWNT-wp} 
\end{figure}

As second calculation route, the energy-dependent and time-dependent
wave-packet spreading
$\ell(E,t)$ is analyzed numerically. The resolution of the time-dependent
Schr\"{o}dinger equation is made by expanding the evolution operator
$e^{-i\hat{\cal H}t}$ on a basis of orthogonal polynomials. This
method has been demonstrated to provide an efficient real-space
computational framework for an order $N$ algorithm~\cite{Roche1999}.
The total length of the MWNTs is finite but several tens of $\mu$m-long
in order to limit boundary effects. For example, the calculations
shown in Figs.~\ref{fig:DWNT-wp} and \ref{fig:TWNT-wp} are performed
with a finite length of about $\unit[24]{\mu m}$, whereas the unit time
step is taken as $T=14\hbar/\gamma_{0}\simeq \unit[3.18]{fs}$.  
Most generally,
the anomalous quantum spreading is driven by a time dependent power
law, $\ell(E,t) = t^{2\eta(E)-1}/v(E)$, whose asymptotic regime gives the
conduction regime. 
When $\ell(E,t)$ tends to some
finite constant at large times, some energy-local elastic mean free path
$\ellel(E)$ can be extrapolated as $\ellel(E) = \lim_{t\to\infty} \ell(E,t)$. In
Fig.~\ref{fig:DWNT-wp} and Fig.~\ref{fig:TWNT-wp}(main frame), the
energy-dependent length scale $\ell(E,t)$ is shown at several chosen
elapsed times $t$, starting from wave packets that are homogeneously
spread in real space with random phase on each lattice point.

For the case of $(9,0)@(10,10)$ (Fig.~\ref{fig:DWNT-wp}), and energies
in the charge neutrality point vicinity, a careful analysis of
propagation over several tens of
microns reveals no deviation from a ballistic-like regime with
$\eta(E)=1$. Conversely for a small energy window in a high density of
states region, a very slowly varying length is observed. Here the
effect of incommensurability becomes more pronounced due to the
enlarged electronic population available for scattering. The result
of the energy-averaged diffusion exponent gives the global exponent
$\eta=0.88$ intermediate between the value $1$ and $1/2$, and in
agreement with prior analysis.

In the case of stronger incommensurate systems, such as the 
$(6,4)@(17,0)@(15,15)$ i-TWNT, the tendency towards energy-dependent
anomalous conduction becomes even more pronounced. First the
region of $\unit[2]{eV}$ around the charge neutrality point remains
almost ballistic, whereas the rest of the electronic spectrum shows a
very slower expansion of wave packet in time. The propagation time runs
from $\unit[0.4]{ps}$ ($t=2000T$) to $\unit[1.5]{ps}$ ($t=6710T$), and
clearly 
$\ell(E,t)$ either shows anomalously slow diffusion, or saturates at very
short times (inset in Fig.~\ref{fig:TWNT-wp}). Whenever the saturation
limit is clearly reached, a
mean free path $\ellel(E)$ for the whole object can be
meaningfully extracted. $\ellel$ is found to be around
$\unit[200]{nm}$ to $\unit[400]{nm}$ for energies in between $\unit[2]{eV}$ to
$\unit[3.5]{eV}$ 

To conclude, we found that the energy-averaged exponent $\eta$ depends
crucially on chirality and number of coupled shells with global anomalous
behavior strongly pronounced in i-DWNT. When looking to
energy-dependent properties, we found that even for i-MWNT exhibiting 
$\eta(E) \sim 1$ (ballistic spreading) close to the charge neutrality point,
some elastic mean free path can be defined in regions of larger
density of states where $\eta(E) \sim 1/2$.

\end{document}